\documentclass[a4paper]{jpconf}
\usepackage{graphicx}
\usepackage{bm}
\usepackage{epsf}
\usepackage{amssymb}
\usepackage{amsmath}
\usepackage{graphicx}
\usepackage{rotating}
\usepackage{epsfig}
\usepackage{psfrag}
\usepackage{amsmath}
\usepackage{hyperref}
\usepackage{subfigure}

\hypersetup{
    bookmarks=true,         
    unicode=false,          
    pdfmenubar=true,        
    pdffitwindow=true,      
    pdftitle={My title},    
    pdfauthor={Author},     
    pdfsubject={Subject},   
    pdfcreator={Creator},   
    pdfproducer={Producer}, 
    pdfkeywords={keywords}, 
    pdfnewwindow=true,      
    colorlinks=true,       
    linkcolor=red,          
    citecolor=green,        
    filecolor=magenta,      
    urlcolor=cyan           
}

\begin{document}
\def\ea{\textit{et al.}}
\def\bF{{\mathbf F}}
\def\ba{{\mathbf a}}
\def\bB{{\mathbf B}}
\def\bE{{\mathbf E}}
\def\bj{\bm{j}}
\def\bJ{{\mathbf J}}
\def\bA{{\mathbf A}}
\def \xy{$x$--$y$ }
\def\bP{{\bf P}}
\def\bK{{\bf K}}
\def\bk{{\bf k}}
\def\bkn{{\bf k}_{0}}
\def\bx{{\bf x}}
\def\bz{{\bf z}}
\def\bR{{\bf R}}
\def\br{{\bf r}}
\def\bq{{\bf q}}
\def\bp{{\bf p}}
\def\bQ{{\bf Q}}
\def\bs{{\bf s}}
\def\bG{{\mathbf G}}
\def\bv{{\bf v}}
\def\b0{{\bf 0}}
\def\la{\langle}
\def\ra{\rangle}
\def\Im{\mathrm {Im}\;}
\def\Re{\mathrm {Re}\;}
\def\beq{\begin{equation}}
\def\eeq{\end{equation}}
\def\bea{\begin{eqnarray}}
\def\eea{\end{eqnarray}}
\def\bdm{\begin{displaymath}}
\def\edm{\end{displaymath}}
\def\bnab{{\bm \nabla}}
\def\Tr{{\mathrm{Tr}}}
\def\sfrac{\textstyle\frac}
\def\Sr{\mathrm{Sr}_2\mathrm{RuO}_4}

\newcommand{\appropto}{\mathrel{\vcenter{
  \offinterlineskip\halign{\hfil$##$\cr
    \propto\cr\noalign{\kern2pt}\sim\cr\noalign{\kern-2pt}}}}}

\title{Anomalous Hall conductivity of clean $\Sr$ at finite temperatures}

\author{Edward~Taylor and Catherine~Kallin}
\address{Department of Physics and Astronomy, McMaster University, Hamilton, Ontario, L8S 4M1, Canada}
\ead{kallin@mcmaster.ca}

\begin{abstract}
Building on previous work, we calculate the temperature- and frequency-dependent {\it anomalous} Hall conductivity for the putative multiband  chiral superconductor $\Sr$ using a simple microscopic two-orbital model without impurities.  A Hall effect arises in this system without the application of an external magnetic field due to the time-reversal-symmetry breaking chiral superconducting state.  The anomalous Hall conductivity is nonzero only when there is more than one superconducting order parameter, involving inter- as well as intra-band Cooper pairing.  We find that such a multiband superconducting state gives rise to a distinctive resonance in the frequency-dependence of the Hall conductivity at a frequency close to the inter-orbital hopping energy scale that describes hopping between Ru $d_{xz}$ and $d_{yz}$ orbitals.  The detection of this feature, robust to temperature and impurity effects in the superconducting phase, would thus constitute compelling evidence in favour of a multiband origin of superconductivity in $\Sr$, with strong superconductivity on the $\alpha$ and $\beta$ bands.  The temperature dependence of the Hall conductivity and Kerr rotation angle are studied within this model at the one-loop approximation.  
\end{abstract}

\section{Introduction}

Since superconductivity was discovered in the layered perovskite $\Sr$ in 1994~\cite{Maeno95}, there has been much speculation as to the nature of the superconducting order parameter.  Almost immediately, it was realized that $\Sr$ might be chiral $p$-wave~\cite{Rice95,Baskran96}, an electronic analogue of the $A$-phase of $^3$He.  However, after more than a decade and a half of intense experimental and theoretical work, basic questions---is the order parameter chiral and if so, where on the three active bands of $\Sr$ does it live?---remain controversial.  (For recent reviews, see Refs.~\cite{Kallin12,Maeno11}, as well as the comprehensive earlier review in Ref.~\cite{Mackenzie03}.)  

If $\Sr$ turns out to be chiral $p$-wave, it would be the only known example so far of this remarkable state in electronic systems.  Chiral superconductivity breaks time-reversal symmetry with order parameters describing Cooper pairs with nonzero magnetic moments.  For chiral $p$-wave  superconductivity, the time-reversal symmetry breaking triplet order parameter is 
\beq \Delta_{\bk} \propto \langle c^{\dagger}_{\bk}c^{\dagger}_{-\bk}\rangle \sim k_x \pm ik_y,\eeq
corresponding to a nonzero angular momentum $\mathbf{L}_z = \pm \hbar$ in the $z$-direction per Cooper pair.  Apart from the intrinsic interest in observing and understanding such an unusual quantum state, the quest to find chiral $p$-wave  superconductors has generated enormous interest since they have been predicted to harbour Majorana bound states~\cite{Read00}, quasiparticles which are their own antiparticles and which may play an important role in quantum computation.  

One important feature of $\Sr$ that can impact the nature of the possibly chiral order parameter is the fact that it is multiband~\cite{Puetter12}.   Three atomic orbitals dominate the electronic structure of $\Sr$ close to the Fermi surface: Ru $d_{xy}$, $d_{xz}$, and $d_{yz}$ orbitals (strongly hybridized with O $p$ orbitals).  The $d_{xz}$ and $d_{yz}$ orbitals hybridize to form the quasi-one-dimensional $\alpha$ and $\beta$ bands, while the $d_{xy}$ orbital predominates the $\gamma$ band, which has an approximately isotropic two-dimensional dispersion.  Spin-orbit coupling leads to some hybridization between the $d_{xz}/d_{yz}$ and $d_{xy}$ orbitals, which otherwise would not hybridize~\cite{Haverkort08}.  Importantly for $\Sr$, the Fermi surface crosses all three of the bands resulting from this orbital structure.  

Where does superconductivity arise amongst these three bands?  The most widely-held viewpoint holds that superconductivity arises primarily on the $\gamma$ band (see, for instance, Refs.~\cite{Agterberg97,Zhitomirsky01,Nomura02,Deguchi04}).  A persistent contrarian position, however, maintains that superconductivity instead arises primarily on the $\alpha$ and $\beta$ bands~\cite{Takimoto00,Kuroki01,Annett02,Raghu10}.  The latter model has one very appealing feature insofar as it may give rise to negligible spontaneous supercurrents at sample edges~\cite{Raghu10}.  Such edge currents have been predicted to arise in one-band chiral $p$-wave superconductors~\cite{Stone04}, but bafflingly, have never been observed~\cite{Kirtley07}.  (For a recent review of experimental probes of the order parameter in $\Sr$, see Ref.~\cite{Kallin09}.) Non-topological chiral superconductivity on the $\alpha$ and $\beta$ bands could explain this.

Amongst experiments purporting to show evidence for time-reversal symmetry breaking in $\Sr$, one of the most direct probes is the Kerr rotation experiment carried out in the group of Kapitulnik at Stanford~\cite{Xia06,Kapitulnik09}.  In this experiment, a near-infrared beam ($\hbar\omega=0.8$eV) is reflected off a sample of $\Sr$, the beam's axis of polarization rotating in the process.  The angle of rotation is called the \emph{Kerr angle}.  It is related to the anomalous Hall conductivity, $\sigma_H$, by~\cite{Argyres55}
\beq \theta_K(\omega) = (4\pi/\omega d)\mathrm{Im}[\sigma_H(\omega)\alpha(\omega)].\label{Kerr}\eeq
Here, $d$ is the interlayer spacing and $\alpha \equiv 1/(n(n^2-1)$ with $n(\omega)$ the complex, frequency-dependent index of refraction.  As we discuss in detail in Sec.~2, a nonzero Hall conductivity indicates that time-reversal symmetry is broken.  

If time-reversal symmetry is broken by chiral superconductivity, spectroscopic knowledge of the frequency-dependent Kerr angle can tell us a lot about the chiral superconducting order parameter in $\Sr$ -- in particular, where it lies in relation to the three active bands of $\Sr$.  (Spectroscopy of Leggett-like collective modes~\cite{Chung12} would also provide considerable insight into this matter.) It is the point of this paper to expand on this idea and suggest ways for future experiments to make further progress in answering questions related to the nature of the order parameter.  

On its own, time-reversal symmetry breaking is not sufficient to have a nonzero Hall conductivity in the absence of an applied magnetic field and hence, a non-zero Kerr angle.  As pointed out by Read and Green~\cite{Read00}, for instance, the Hall conductivity of a translationally-invariant chiral $p$-wave  superconductor is identically zero~\cite{Roy08}.  Since the publication of  Kapitulnik's group's data in 2006~\cite{Xia06}, two credible mechanisms have emerged in which chiral $p$-wave  superconductivity in $\Sr$ can give rise to a nonzero anomalous Hall conductivity.  First, a purely extrinsic mechanism due to impurity scattering was suggested by Goryo~\cite{Goryo08} and Lutchyn \textit{et al.}~\cite{Lutchyn09}.  Subsequently, an \emph{intrinsic} mechanism was identified by us~\cite{Taylor12} as well as Wysoki\'{n}ski~\textit{et al.}~\cite{Wysokinski12}.  (A previously identified intrinsic contribution from chiral p-wave collective modes~\cite{Yip92} is estimated to be several orders of magnitude too small to explain the experiments~\cite{Xia06} and will not be discussed here.)

Very usefully for the questions identified above, the intrinsic mechanism gives rise to a nonzero anomalous Hall conductivity if and only if there are multiple chiral order parameters spanning multiple optically active bands.  For $\Sr$, this means that there \emph{must} be superconductivity on the $\alpha$ and $\beta$ bands in order for the intrinsic $\sigma_H$ to be nonzero.  In contrast, the extrinsic mechanism contributes as well when chiral superconductivity occurs on the $\gamma$ band.  Thus, if it was found that the intrinsic mechanism dominated the Hall conductivity, one could reliably conclude that the origin of superconductivity in $\Sr$  lay in the $\alpha$ and $\beta$ bands and not the $\gamma$ band.  

In this paper, we suggest two ways in which experiments can deduce the relative importance of the intrinsic and extrinsic mechanisms.  First, the intrinsic mechanism produces a very strong spectral feature in the frequency-dependent anomalous Hall conductivity (and Kerr angle) close to twice the inter-orbital $d_{xz}-d_{yz}$ hopping energy, near $0.1$eV.  The detection of this feature, robust to temperature and impurity effects in clean samples of $\Sr$, would provide nearly unequivocal evidence in our opinion for strong superconductivity on the $\alpha$ and $\beta$ bands.  Second, by measuring the dependence of $\sigma_H$ on the impurity concentration, knowing the dependence of the superconducting gap on this quantity as well, one could estimate the relative roles of the intrinsic and extrinsic contributions.

\section{Chiral two-band superconductor}

Following our earlier work~\cite{Taylor12}, we analyze a two-orbital model in order to make clear the physics of the anomalous Hall conductivity in a multiband superconductor at finite temperatures. We will use this model later to describe the physics of $\Sr$ since, although three orbitals are needed to account for the Fermi surface properties~\cite{Mackenzie03}, only two orbitals--the Ru $d_{xz}$ and $d_{yz}$ ones--are directly relevant for the Hall conductivity.  

We consider the following BCS pairing Hamiltonian describing superconductivity in a two-orbital system:
\beq H\! =\! \sum_{\bk}\!\left(\!\begin{array}{cc} c^{\dagger}_{\bk1} & c^{\dagger}_{\bk2}\end{array}\!\right)\left(\!\begin{array}{cc}
\xi_1(\bk)\! & \!\epsilon_{12}(\bk) \\ \epsilon_{12}(\bk) \!& \!\xi_2(\bk)
\end{array}
\right)\left(\begin{array}{c} 
c_{\bk 1} \\ c_{\bk 2}
\end{array}\right) + \sum_{\alpha,\beta}\sum_{\bk,\bk'}V_{\alpha,\beta}(\bk,\bk')c^{\dagger}_{-\bk \alpha}c^{\dagger}_{\bk\beta}c_{\bk'\beta}c_{-\bk'\alpha}.\label{H}\eeq
Here, $\xi_{1(2)}\equiv \epsilon_{1(2)}-\mu_{1(2)}$ is the dispersion for the Bloch states constructed from the 1(2) orbital, $\epsilon_{12}$ is the inter-orbital coupling, and $V_{\alpha,\beta}(\bk,\bk')$ describes the pairing interaction, allowing for the possibility of inter-orbital ($\alpha=1,\beta=2$ and $\alpha=2,\beta=1$) as well as intra-orbital ($\alpha=\beta=1,2$) pairing.  In the discussion to follow, we use $\Delta_{11}$ and $\Delta_{22}$ to denote the two intra-orbital order parameters while $\Delta_{12}=\Delta_{21}$ is the inter-orbital order parameter.   Spin labels have been suppressed.

In the  basis set by the spinor, $ \hat{\Psi}^{\dagger}_{\bk} = (c^{\dagger}_{\bk 1},c_{-\bk 1},c^{\dagger}_{\bk 2},c_{-\bk 2})$, the inverse mean-field  $4\times 4$ Green's function for this model is 
\begin{align} \bG^{-1}_0(\bk,\omega_n) =\left(\!\begin{array}{cc}
\!i\omega_n \!-\! \xi_{1}\hat{\tau}_3\!+\! \Delta^{\prime}_{11}\hat{\tau}_1\!-\!\Delta^{\prime\prime}_{11}\hat{\tau}_2 &\! -\epsilon_{12}\hat{\tau}_3 \!+\! \Delta^{\prime}_{12}\hat{\tau}_1\!-\!\Delta^{\prime\prime}_{12}\hat{\tau}_2 \\
\!-\epsilon_{12}\hat{\tau}_3 \!+\! \Delta^{\prime}_{12}\hat{\tau}_1\!-\!\Delta^{\prime\prime}_{12}\hat{\tau}_2&\! i\omega_n \!-\! \xi_{2}\hat{\tau}_3\! +\! \Delta^{\prime}_{22}\hat{\tau}_1\!-\!\Delta^{\prime\prime}_{22}\hat{\tau}_2
\end{array}\!\right). \label{G0}\end{align}
Here, $\hat{\tau}_l$ are the usual $2\times 2$ Pauli matrices,
$\Delta^{\prime}_{a}$ ($\Delta^{\prime\prime}_{a}$) is the real (imaginary) part of the intra- and inter-orbital order parameters, and $\omega_n$ is a Fermi Matsubara frequency.  The two branches of the BCS quasiparticle spectrum, $E_{-}$ and $E_{+}$, found from the solution of $\mathrm{det}\bG^{-1}_0(\bk,\omega_n) = (\omega^2_n+E^2_{-})(\omega^2_n+E^2_{+})$, are
\begin{align} E^2_{\pm} &= \frac{1}{2}\Big\{|\Delta_{11}|^2+|\Delta_{22}|^2+2|\Delta_{12}|^2+2\epsilon^2_{12}+\xi^2_{1}+\xi^2_{2}\pm \big[(|\Delta_{11}|^2+|\Delta_{22}|^2+2|\Delta_{12}|^2+2\epsilon^2_{12}+\xi^2_{1}\nonumber\\&+\xi^2_{2})^2 -4\big(|\Delta_{22}|^2\xi^2_{1} +|\Delta_{11}|^2\xi^2_2 + |\Delta_{11}|^2|\Delta_{22}|^2 -4\epsilon_{12}\xi_1\mathrm{Re}[\Delta_{12}\Delta^{*}_{22}]-4\epsilon_{12}\xi_2\mathrm{Re}[\Delta_{12}\Delta^{*}_{11}]\nonumber\\&+2\epsilon^{2}_{12}\mathrm{Re}[\Delta_{11}\Delta^{*}_{22}]-2\mathrm{Re}[\Delta_{11}\Delta_{22}(\Delta^{*}_{12})^2]+ (\epsilon^2_{12}-\xi_{1}\xi_2)^2\big)\big]^{1/2}\Big\} .\end{align}

In the situation (relevant for $\Sr$) that the Fermi surface crosses the two sheets formed by the bare 1 and 2  orbital dispersions and $\mu_1\simeq\mu_2\equiv \mu \gg |\Delta_{ij}|$ $\forall i,j$, the minimum in the \emph{sum} of the two Bogoliubov quasiparticle bands, $E_-+E_+$, is close to twice the inter-orbital hopping energy:
\beq \mathrm{min}(E_-+E_+) \sim 2\epsilon_{12}(k_x\simeq k_y \simeq |\bk_0|).\label{QPmin}\eeq
Here, $\bk_0$ are the wavevectors at which the bare dispersions $\xi_1$, $\xi_2$ both vanish; i.e., where the Fermi surface crosses the 1 and 2 sheets.   
In this case, as we will see in the following, the inter-orbital hopping energy scale plays a crucial role in determining the frequency dependence of the anomalous, intrinsic Hall conductivity in a multiband chiral superconductor, with multiple order parameters spanning the bands.

\section{Intrinsic and anomalous Hall conductivity}

\subsection{Formalism}
The optical Hall conductivity $\sigma_H(\omega)$ is defined in terms of the antisymmetric part of the $\hat{J}_x$-$\hat{J}_y$ current correlator $\pi_{xy}(\bq,\omega)$ by
\beq \sigma_H(\omega) \equiv -\frac{1}{2i\omega}\lim_{\bq\to 0}\left[\pi_{xy}(\bq,\omega) - \pi_{yx}(\bq,\omega)\right].\label{sigmaH}\eeq
The total current operator in the $i$ direction is given by~\cite{note} $\hat{J}_{i} =e\sum_{\bk}\mathrm{tr}\hat{\Psi}^{\dagger}_{\bk}\hat{\mathbf{v}}(\bk)_{i}\hat{\Psi}_{\bk}$, where
\beq \hat{\mathbf{v}}_{i} (\bk)= \left(\begin{array}{cc}
v_{i,11}(\bk)\hat{1}_2  & v_{i,12}(\bk) \hat{1}_2 \\
v_{i,12}(\bk) \hat{1}_2 & v_{i,22}(\bk)\hat{1}_2
\end{array}\right)\label{V}\eeq
is the $4\times 4$ bare current vertex ($\hat{1}_2$ is the $2\times 2$ identity matrix).  In the orbital basis, $v_{i,aa}(\bk) = \partial_{k_{i}}\epsilon_a$ and  $v_{i,12}(\bk) = \partial_{k_{i}}\epsilon_{12}$.

When there are multiple orbitals present, it is possible to have a nonzero anomalous Hall conductivity evaluating the current correlator at the one-loop level, and so in the following we ignore vertex corrections:
\beq \pi_{xy}(\b0,\!\nu_m) \!=\!e^2T\!\!\sum_{\bk,\omega_n}\!\mathrm{tr}[\hat{\mathbf{v}}_x(\bk) \bG_0(\bk,\!\omega_n)\hat{\mathbf{v}}_y(\bk)\bG_0(\bk,\!\omega_n+\nu_m)].\label{pi0}\eeq
Here, $\nu_m$ is a Bose Matsubara frequency. In the case of a single orbital (or multiple uncoupled orbitals), $\hat{\mathbf{v}}_{\sigma}$ is purely diagonal and commutes with $\bG_0$.  Consequently, $\pi_{xy}$ equals $\pi_{yx}$, and the one-loop value for the Hall conductivity is zero, irrespective of details such as band anisotropy, pairing symmetry, or self-energy corrections~\cite{Lutchyn09,Taylor12}.   Broken time-reversal and translational symmetries are necessary but not sufficient conditions for a nonzero anomalous Hall conductivity at this level.  It follows that for superconductivity on a single orbital, vertex corrections are crucial to having a nonzero anomalous Hall conductivity.  Goryo~\cite{Goryo08} and Lutchyn \textit{et al.}~\cite{Lutchyn09} considered impurity-scattering vertex corrections for a model of superconductivity in $\Sr$ assuming superconductivity takes place predominantly on the $\gamma$ band (or $d_{xy}$ orbital).   

For a multiorbital superconductor, on the other hand, the one-loop contribution (\ref{pi0}) can be nonzero, and if so, should provide a major contribution to the anomalous Hall effect in a clean superconductor. This contribution is straightforwardly evaluated by analytically continuing  to real frequencies, $i\nu_m\to \omega+i0^+$, to obtain the real and imaginary parts of the anomalous Hall conductivity (\ref{sigmaH}).  The complex conductivity is given by
\begin{align} \sigma_H&(\omega) = 2e^2\sum_{\bk}\left(\delta\bv_{21}\times\bv_{12}\right)_z\left[\epsilon_{12}\mathrm{Im}(\Delta^*_{11}\Delta_{22})+\xi_1\mathrm{Im}(\Delta^*_{22}\Delta_{12})-\xi_2\mathrm{Im}(\Delta^*_{11}\Delta_{12})\right]\times\nonumber\\& \frac{1}{E_{+}E_{-}}\left\{\frac{1\!-\!f(E_{+})\!-\!f(E_{-})}{(E_{-}\!+\!E_{+})[(E_{-}\!+\!E_{+})^2\!-\!(\omega+i\eta)^2]} + \frac{f(E_{+})\!-\!f(E_{-})}{(E_{+}\!-\!E_{-})[(E_{+}\!-\!E_{-})^2\!-\!(\omega+i\eta )^2]}\right\} . \label{sigmareal}\end{align}
Here, $\bv_{ab}\equiv (v_{x,ab},v_{y,ab})$, and $\delta\bv_{21}\equiv \bv_{22}-\bv_{11}$. $\eta= 0^{+}$ is an infinitesimal in the clean limit.  Otherwise, we can make a crude estimate of the effects of impurity scattering by using a finite $\eta\equiv 1/\tau$, where $1/\tau$ is the impurity scattering rate.  For unconventional $p$-wave superconductivity, superconductivity is easily destroyed by impurity scattering and one expects to be in the clean limit, $|\Delta_{ij}|\tau \gg 1\Rightarrow \eta\ll |\Delta_{ij}|$, for samples with the highest $T_c$.

\subsection{Generic features of the Hall conductivity in a multiband chiral superconductor}

Before discussing the implications of Eq.~(\ref{sigmareal}) for $\Sr$, we point out some generic features relevant for any multiorbital chiral superconductor.  At the one-loop level, due to the cancellation between the two terms in Eq.~(\ref{sigmaH}), the only contributions to a nonzero $\sigma_H$ arise from time-reversal symmetry-breaking inter-orbital transitions.  The factor $\left(\delta\bv_{21}\times\bv_{12}\right)_z$ in Eq.~(\ref{sigmareal}) indicates that one photon vertex involves an electron transition between different orbitals, while the other vertex involves an intra-orbital transition.  The two terms in the curly brackets of Eq.~(\ref{sigmareal}) indicate that, in the quasiparticle basis, only processes which create or destroy two quasiparticles on {\it different} branches (first term) or which scatter a quasiparticle from one branch to the other (second term) contribute.  The rest of Eq.~(\ref{sigmareal}), the terms in square brackets together with energy denominators, are the form factors, or coherence factors, that relate the quasiparticle states to the electron orbital states.  In passing from Eq.~(\ref{pi0}) to Eq.~(\ref{sigmareal}), almost all such terms cancel, leaving only two very specific terms which carry the signature of the broken time reversal symmetry of the superconducting state.   The relevant terms contain a coupling between the orbitals, either through the hopping, $\epsilon_{12}$, or through intra-orbital pairing, $\Delta_{12}$, that is essential to connect an inter-orbital vertex with an intra-orbital vertex.  The signature of broken time reversal symmetry comes into these terms through a product of different intra- and inter-orbital superconducting order parameters.  A nonzero \emph{relative} phase of these order parameters,
\beq \Delta\phi_{11,22}\equiv \phi_{11}-\phi_{22}\equiv \mathrm{Im}(\Delta^{*}_{11}\Delta_{22})/|\Delta_{11}||\Delta_{22}|,\eeq
 for example, then gives rise to a nonzero anomalous Hall conductivity.  Consequently, the existence of multiple order parameters with different complex phases spanning multiple orbitals is a necessary condition to have a nonzero anomalous Hall conductivity.  

Beyond the fact that multiple orbitals are needed in order to have a nonzero anomalous Hall conductivity at the one-loop level (i.e., no vertex corrections), Eq.~(\ref{sigmareal}) shows that inter-\emph{band} Cooper pairing is also a necessary condition.  Although this expression shows the Hall conductivity with respect to orbital degrees of freedom, it is straightforward to apply a unitary transformation to represent this in terms of band degrees of freedom.  The second line of Eq.~(\ref{sigmareal}) remains the same since it is expressed in terms of the Bogoliubov quasiparticle basis.  On the other hand, the velocities $\delta\bv_{21}$ and $\bv_{12}$ will be different and the term in square brackets involving the coupling $\epsilon_{12}$ will vanish since the bands are, by definition, uncoupled.  This leaves the remaining terms in square brackets which, in the band basis, describe inter-band pairing, $\Delta_{\alpha\beta}\neq 0$ where $\alpha\neq \beta$ are the band indices.  Thus, inter-band pairing is needed to have a nonzero anomalous Hall conductivity in a chiral superconductor at the one loop level.  Moreover, as in the orbital picture, the inter-band order parameter cannot have the same phase as at least one of the intra-band order parameters.  

At zero temperature, only the first term in curly brackets in Eq.~(\ref{sigmareal})  contributes to the Hall conductivity; it corresponds to a process wherein a photon destroys a Cooper pair and creates two quasiparticles.  The imaginary part of $\sigma_H$, which is the absorptive part, turns on at the minimum of $E_-+E_+$.  As discussed above, this minimum is set by the inter-orbital coupling, $\epsilon_{12}$, and, in general, is expected to be a much larger energy scale than the BCS  quasiparticle gap $|\Delta_{ij}|$.  The zero temperature anomalous Hall conductivity was studied in Ref.~\cite{Taylor12} and the absorptive part exhibits sharp features due to the van Hove singularites in the quasiparticle density of states near the minimum of $E_-+E_+$. 

There are two effects of finite temperature: (1) to modify the contribution from the first term in curly brackets and (2) the new contribution represented by the second term in the curly brackets.   If $\Delta_{ij}\ll \epsilon_{12}(\bk_0)$, the primary effect of (1) is to replace the zero temperature values of the $\Delta$'s with their finite temperature values.  In particular, the sharp features emerging due to the van Hove singularities near the minimum of $E_-+E_+$  (see Figs.~\ref{SigmaOmegaTeq0fig} and \ref{SigmaOmegaTeq1pt4fig}) will not be appreciably smoothed out at nonzero temperatures.  This is because, relative to $\epsilon_{12}(\bk_0)$, the temperature, $T\leq T_c\sim \Delta$, is always very small in the superconducting phase where the anomalous Hall effect arises.  For this same reason, contribution (2), due to quasiparticle scattering, is also very small.  This contribution vanishes at $T=0$, as well as at $T=T_c$, and is largest at temperatures comparable to but less than $T_c$.    For most frequencies, this term is smaller than contribution (1) by a factor of order $\Delta/\epsilon_{12}({\bf k}_0)$ because the Fermi factors restrict the momentum integral to a very narrow region where $E_1\sim\Delta$.  Consequently,  at most frequencies, and in particular at the high frequency relevant to current experiments, the anomalous Hall conductivity is dominated by the first term in the curly brackets and the main effect of finite temperature is simply to replace the zero temperature  $\Delta$ with $\Delta(T)$.

As noted above, the intrinsic Hall conductivity arises from inter-orbital transitions when there are order parameters of different phases living on the orbitals.  The van Hove singularities represent a resonance condition for these transitions.  In this way, the detection of a strong increase in the Hall conductivity at frequencies on the order of the inter-orbital hopping energy would provide strong evidence in favour of multiband/multiorbital Cooper pairing in a chiral superconductor, such as has been proposed in $\Sr$~\cite{Takimoto00,Kuroki01,Annett02,Raghu10}.  In contrast, features in the Hall conductivity arising at frequencies on the order of the gap, such as impurity scattering contributions which yield a nonzero Hall conductivity in a single band system~\cite{Goryo08,Lutchyn09}, are comparatively insensitive to whether or not superconductivity is a multiband phenomenon.

\section{Optical Hall conductivity in $\Sr$} 
Having used a two-orbital model to elucidate the basic physics of the anomalous intrinsic Hall effect in a multiorbital chiral superconductor, we now discuss the implications for $\Sr$.  Although $\Sr$ is a three-band system, in calculating the intrinsic anomalous Hall conductivity, $\sigma_H$, an important simplification arises because, to a very good approximation, the $x$ and $y$ components of the current operator do not couple $d_{xy}$ to the other two orbitals and, hence, only $d_{xz}-d_{yz}$ inter-orbital transitions contribute to its $\sigma_H$.  Consequently, the conclusions reached from our two orbital model are still valid.  In particular, the intrinsic, anomalous Hall effect in $\Sr$ is only nonzero when there is complex inter-band pairing and this pairing must  involve the $d_{xz}$ and $d_{yz}$ orbitals to a significant extent.   Higher order spin-orbit coupling, of order $\lambda^2$ (where $\lambda$ is the spin-orbit coupling parameter; see e.g. Ref.~\cite{Rozbicki11}), and interlayer hopping involving $d_{xy}$ orbitals, will bring in the $d_{xy}$ orbital, but these are are sufficiently small that they can be ignored.  Consequently, the $d_{xy}$ orbital only plays a passive role in determining the Hall conductivity.  In particular, the terms outside the curly brackets in Eq.~(\ref{sigmareal}) would be unchanged if we used a three-orbital model of $\Sr$, but neglected interlayer hopping and higher order spin-orbit effects.  Only  the quasiparticle dispersions would be modified and would renormalize our estimate of the Hall conductivity somewhat.  Thus, we simply ignore the $d_{xy}$ orbital and use our two-orbital result for the Hall conductivity, (\ref{sigmareal}), as this gives a reasonable order of magnitude estimate.

Building on our previous work~\cite{Taylor12} where we neglected next-nearest neighbour intra-orbital hopping, we now include this effect, taking $\epsilon_1 =  -2t\cos(k_x)-2t^{\prime\prime}\cos(k_y)$, $\epsilon_2  =-2t\cos(k_y)-2t^{\prime\prime}\cos(k_x)$, and $\epsilon_{12}=2t^{\prime}\sin(k_x)\sin(k_y)$.  We use parameters from a recent LDA study of the band structure of $\Sr$~\cite{Rozbicki11}:  $t= 0.4$eV, $t^{\prime} = 0.1t$, and $t^{\prime\prime}=0.125 t$.   We also fix the chemical potential to be $\mu = t$, consistent with this study.    
Following Ref.~\cite{Raghu10}, we assume purely intra-orbital pairing on the $d_{xz}$ and $d_{yz}$ orbitals.  As a simple ansatz, we use \beq\Delta_{11}(\bk,T) = \Delta_0(T)\sin(k_x)\cos(k_y),\; \Delta_{22}(\bk,T) = i\Delta_0(T)\sin(k_y)\cos(k_x),\; \Delta_{12}(\bk,T)=0.\label{ansatz}\eeq
 \begin{figure}
\begin{center}
\includegraphics[width=0.45\textwidth]{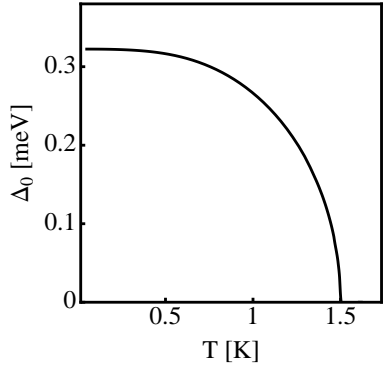}
\caption{Temperature dependence of the order parameter amplitude, $\Delta_0(T)$, defined in Eq.~(\ref{delta0}).
}
\label{gapfig}
\end{center}
\end{figure}
Here, $\Delta_0(T)$ is the purely real temperature-dependent amplitude of the order parameters, assumed to be equal for both.  Crucially in order to obtain a nonzero Hall conductivity, the two order parameters have different phases, meaning that $\mathrm{Im}(\Delta^{*}_{11}\Delta_{22})$ and hence, $\sigma_H$, is nonzero.

Mineev~\cite{Mineev12} has argued that a model of the form shown in Eq.~(\ref{ansatz}), based on a quasi-1D approach, is not
energetically stable in the case relevant to $\Sr$, where
$t^{\prime}\gg \Delta_0$.  However, since $t^{\prime}\ll t$, $t^{\prime}$ only affects a very small region
of the Fermi surface, while $\Delta_0$ operates over the entire
Fermi surface.  As explained by Chung \textit{et al.}~\cite{Chung12}, this allows the
quasi-1D approach to remain valid even for $t^{\prime}\gg\Delta_0$.  While we
use a somewhat simplified model, more realistic details, such
as allowing different pairing amplitudes on the $\alpha$ and
$\beta$ bands would not change our results in any significant
way.  \emph{The key property of the model is that it includes
\underline{interband pairing}.}  As stressed in Ref.~\cite{Taylor12}, this is a
necessary condition for a non-zero intrinsic Hall effect,
a point also emphasized by Mineev~\cite{Mineev12}.

To see the structure of the purely intraorbital order parameter, Eq.~(\ref{ansatz}) in the band basis, we diagonalize the single-particle contribution to the Hamiltonian, Eq.~(\ref{H}), into the $\alpha$ and $\beta$ bands.  In this basis, there are three order parameters: two intra-band,
\beq \Delta_{\alpha\alpha} = \frac{1}{(\xi_+-\xi_1)^2+\epsilon^2_{12}}\left[\epsilon^2_{12}\Delta_{11} + (\xi_+-\xi_1)^2\Delta_{22}\right],\eeq
\beq \Delta_{\beta\beta} = \frac{1}{(\xi_--\xi_1)^2+\epsilon^2_{12}}\left[\epsilon^2_{12}\Delta_{11} + (\xi_--\xi_1)^2\Delta_{22}\right],\eeq
and one inter-band,
\beq \Delta_{\alpha\beta}= \Delta_{\beta\alpha} = \frac{1}{\sqrt{(\xi_+-\xi_1)^2+\epsilon^2_{12}}\sqrt{(\xi_--\xi_1)^2+\epsilon^2_{12}}}\left[\epsilon^2_{12}\Delta_{11} + (\xi_+-\xi_1)(\xi_--\xi_1)\Delta_{22}\right],\eeq
order parameters.  
 

As expected, the inter- and intra-band order parameters have different phases and hence, the Hall conductivity will be nonzero, as we already inferred from the fact that the phases of the order parameters in the orbital basis were also different.  It is interesting to note that in the ``hot zone" $\xi_1\simeq \xi_2$ (with $k_x \simeq k_y\simeq |\bk_0|$) which delineates the regions of the Brillouin zone that dominate the conductivity integral in Eq.~(\ref{sigmareal}), these order parameters have approximately opposite chirality:
\beq \Delta_{\alpha\alpha}\simeq \Delta_{\beta\beta}\simeq \left[\Delta_{11}+\Delta_{22}\right]/2\eeq
and
\beq \Delta_{\alpha\beta}\simeq \left[\Delta_{11}-\Delta_{22}\right]/2.\eeq

Turning now to the task of evaluating numerically the Hall conductivity, we assume a separable potential, $V_{\alpha\alpha}(\bk,\bk') = gf^{(\alpha)}_{\bk}f^{(\alpha)}_{\bk'}$ where $f^{(1/2)}_{\bk}=\sin(k_{x/y})\cos(k_{y/x})$, and use this to determine the amplitude $\Delta_0(T)$ of the order parameter.  The resulting BCS gap equation is
\beq \frac{\Delta_0(T)}{g} =T\sum_{\bk,\omega_n}\sin k_x \cos k_y G_{0,12}(\bk,\omega_n).\label{delta0}\eeq
Here, $G_{0,ij}$ is the $(i,j)$ element of the $4\times 4$ matrix Green's function, the inverse of which is given by Eq.~(\ref{G0}).  
We choose the dimensionless coupling constant (restoring the crystalline spacing $a$ but keeping $k_B=\hbar=1$) $t\pi^2a^2/g$ to be $\simeq 5.59$ in order to have a critical temperature $T_c = 1.5K$~\cite{Xia06}, appropriate for ultraclean $\Sr$ samples.   
We plot $\Delta_0(T)$ in Fig.~\ref{gapfig}.  Note that the $T=0$ value, $\Delta_0(0)\simeq 0.32$meV, is larger than the canonical weak-coupling $s$-wave BCS-limiting value, $\Delta_{0,BCS}(0)\simeq 1.76T_c\simeq 0.23$meV, because of anisotropy.

\begin{figure}[h]
\begin{minipage}[t]{16pc}
\includegraphics[width=16pc]{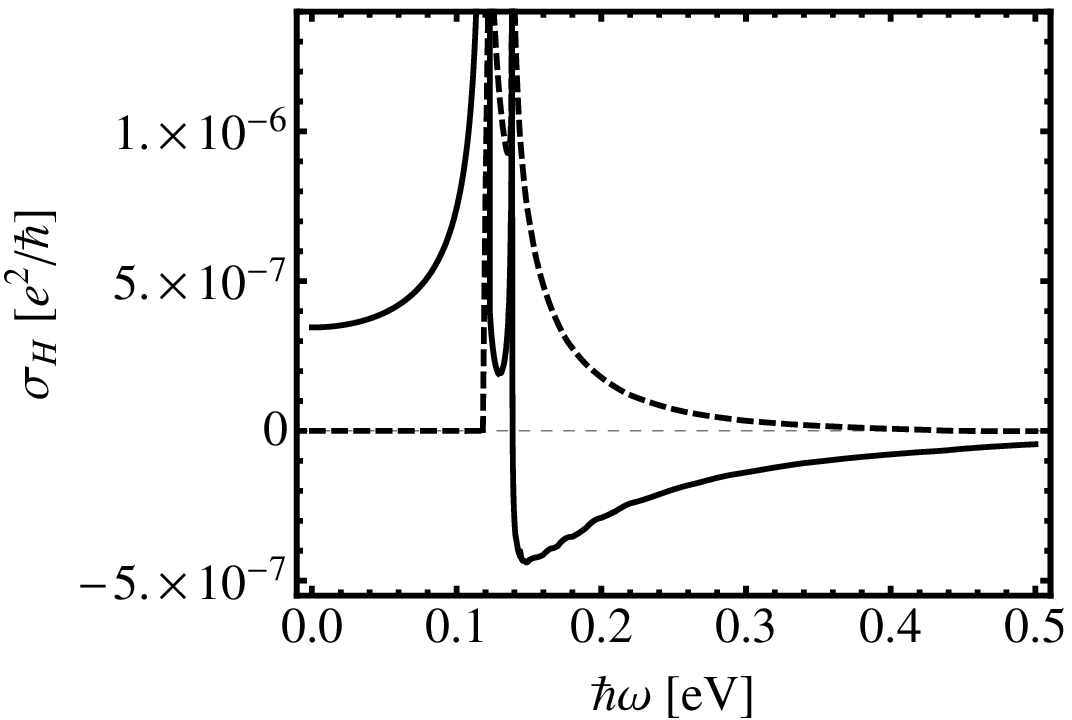}	
 \caption{Frequency dependency of the real (solid line) and imaginary (dashed line) parts of the Hall conductivity at $T=0$.  In all figures, the model parameters are as given in the text surrounding Eqs.~(\ref{ansatz})  and (\ref{delta0}).}
\label{SigmaOmegaTeq0fig}
\end{minipage}
\hspace{1pc}
\begin{minipage}[t]{16pc}
\includegraphics[width=16pc]{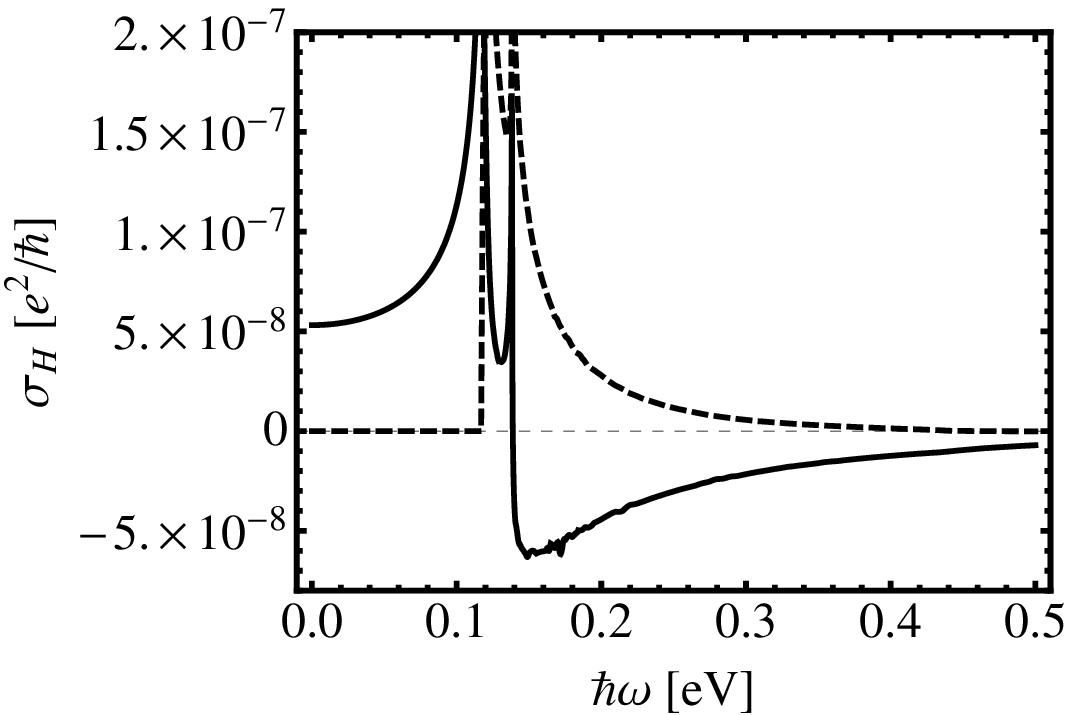}
\caption{Frequency dependence of the real and imaginary parts of the Hall conductivity at $T=1.4K (\sim 0.93T_c)$. Note the different scale as compared to Fig.~\ref{SigmaOmegaTeq0fig}.}
	\label{SigmaOmegaTeq1pt4fig}
\end{minipage}
\end{figure}

\begin{figure}[h]
\begin{minipage}[t]{16pc}
\includegraphics[width=16pc]{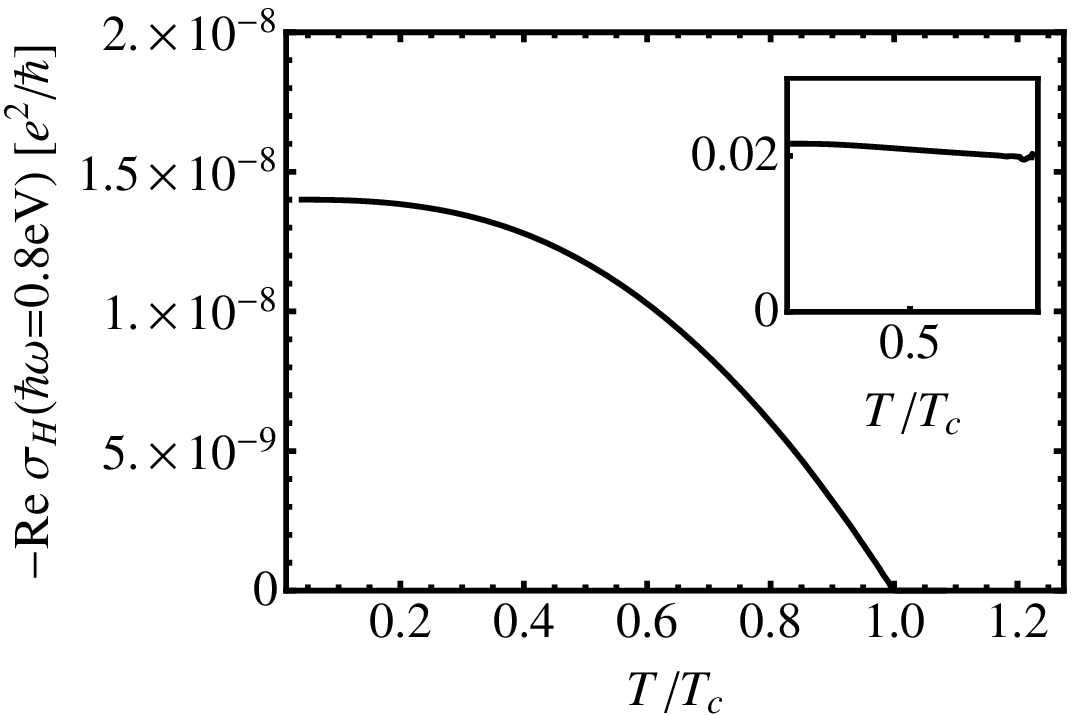}	
 \caption{Temperature dependence of the real part of the Hall conductivity at $\hbar\omega=0.8$eV.  Inset: Temperature dependence of the real part of the Hall conductivity at $\hbar\omega=0.8$eV divided by $(\Delta_0(T)/t)^2$.}
\label{ReSigmaTfig}
\end{minipage}
\hspace{1pc}
\begin{minipage}[t]{16pc}
\includegraphics[width=16pc]{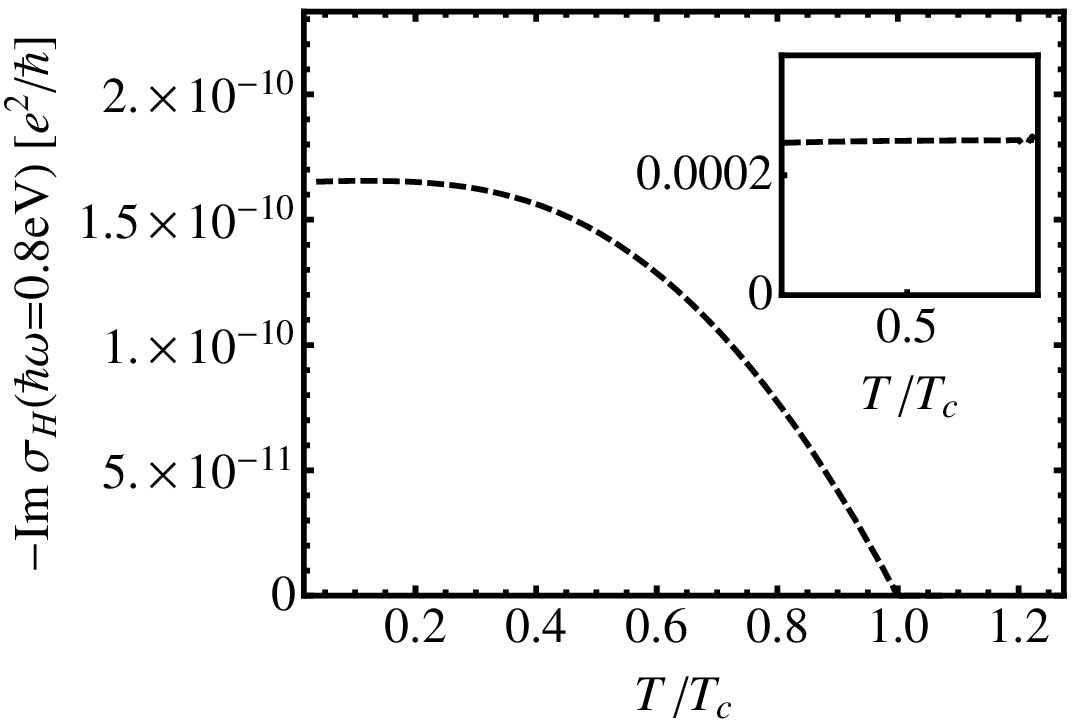}
\caption{Temperature dependence of the imaginary part of the Hall conductivity at $\hbar\omega=0.8$eV. Inset: Temperature dependence of the imaginary part of the Hall conductivity at $\hbar\omega=0.8$eV divided by $(\Delta_0(T)/t)^2$.}
	\label{ImSigmaTfig}
\end{minipage}
\end{figure}

In Figs.~\ref{SigmaOmegaTeq0fig} and \ref{SigmaOmegaTeq1pt4fig}, we make use of the calculated $\Delta_0(T)$ in Eq.~(\ref{sigmareal}) to plot the frequency-dependent Hall conductivity at $T=0$ and $T=1.4K$, respectively in the ultraclean limit, $\eta = 10^{-6}t$.  There is very little difference between the two plots apart from an overall scale factor: $\sigma_H(\omega,T)\simeq (\Delta_0(T)/\Delta_0(0))^2\sigma_H(\omega,0)$.  This can be seen explicitly in the inset of Figs.~\ref{ReSigmaTfig} and \ref{ImSigmaTfig} which show the temperature dependence of the real and imaginary parts of the Hall conductivity (at $\hbar\omega=0.8$eV), respectively, divided  $(\Delta_0(T)/t)^2$.  One sees that this quantity is essentially independent of temperature.  This result,
\beq \sigma_H(\omega,T) \appropto \Delta^2_0(T)\;\;\mathrm{for}\;\;\omega \nsim 2t^{\prime},\label{propid}\eeq
where the proportionality factor is independent of temperature, holds for all frequencies except those very close to the van Hove singularity in the vicinity of $\omega \sim 2t^{\prime}$.  In the main panels of Figs.~\ref{ReSigmaTfig} and \ref{ImSigmaTfig}, we plot the temperature dependence of the real and imaginary parts of the conductivity.

The fact that the Hall conductivity is approximately proportional to the square of the temperature-dependent order parameter (at least away from the van Hove singularities) accounts for its smallness.  Dividing the calculated Hall conductivity by the relevant dimensionless energy scale $x\equiv (\Delta_0(0)/t)^2\sim 6\times 10^{-7}$, we see that the zero-frequency and temperature Hall conductivity $\sigma_H(0)/x \sim  e^2/\hbar$ is approximately unity (in natural units of $e^2/\hbar$).  It is only the smallness of the order parameter in $\Sr$ that makes the anomalous Hall conductivity as small as it is.  

Equation~(\ref{propid}) also suggests that useful information about the origin of the anomalous Hall effect in $\Sr$ can come from looking at the impurity concentration dependence of the Hall conductivity in $\Sr$.  $T_c$ in non-$s$-wave superconductors is strongly suppressed by non-magnetic impurities~\cite{Balian63}.  In $\Sr$, this dependence has been studied in the group of Mackenzie~\cite{Mackenzie98a,Mackenzie98b}.  Assuming that $\Delta_{0}(T\sim 0) \propto T_c$, measurement of the impurity concentration dependence of the Hall conductivity (or Kerr angle, as discussed below) in conjunction with the impurity concentration dependence of $T_c$ would help reveal the relative importance of the intrinsic and extrinsic contributions to $\sigma_H$.

Figures~\ref{SigmaOmegaTeq0fig} and \ref{SigmaOmegaTeq1pt4fig} also clearly exhibit the sharp features anticipated in Sec.~III, arising from the van Hove  singularities near  min$[E_{-}(\bk)+E_{+}(-\bk)]\sim 2t^{\prime}\gg \Delta_0$.  As discussed there, these are not at all washed out at finite temperatures since, relative to this energy scale ($t^{\prime}=40$meV), the temperature is very small in the superconducting phase ($T_c\sim 0.13$meV).  We have also checked the effect of adding a small amount of impurity scattering, taking $\eta = 10^{-4}t$ (corresponding to $\Delta_0\tau\sim 10$) and, unsurprisingly given this small broadening (recall that $\Delta_0 \ll t^{\prime}$), find very little difference from the results shown in Figs.~\ref{SigmaOmegaTeq0fig} and \ref{SigmaOmegaTeq1pt4fig}.  Notably, the resonance features remain intact.  If measured, these strong features, arising as they do from inter-orbital transitions involving the creation of BCS quasiparticles from multiple complex order parameters spanning these orbitals, would constitute strong evidence for multiband chiral superconductivity.  

\begin{figure}[h]
\begin{minipage}[t]{16pc}
\includegraphics[width=16pc]{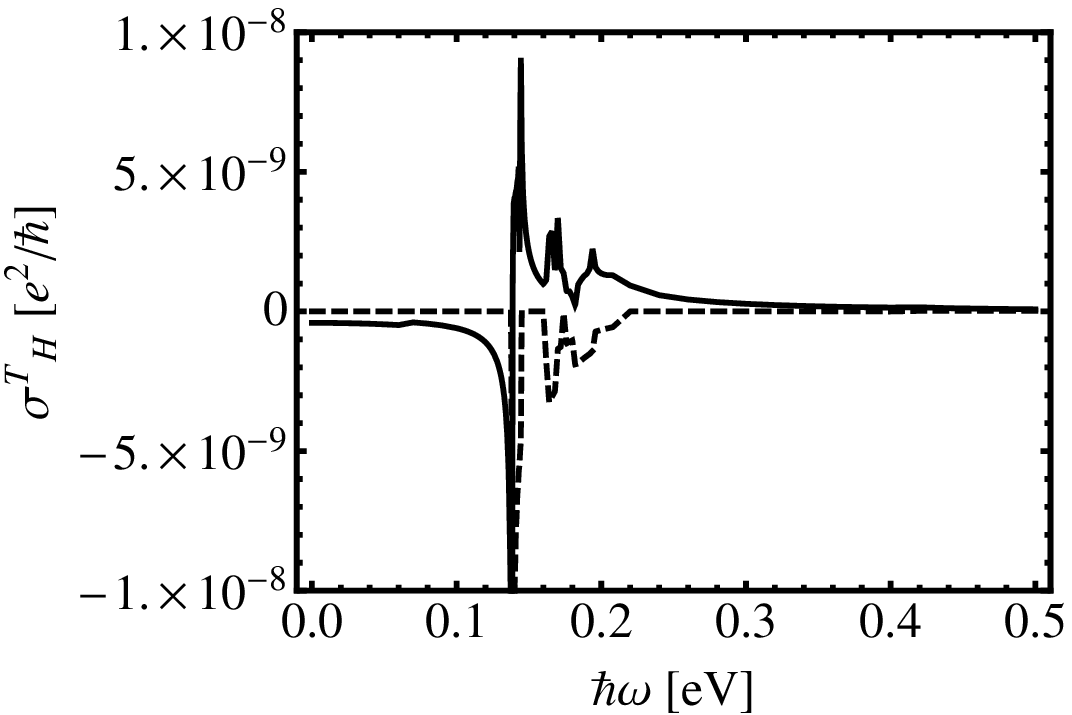}
\caption{Frequency dependence of only the quasiparticle scattering contribution, $\sigma^{T}_H$, to the Hall conductivity, coming from the second term in curly brackets in Eq.~(\ref{sigmareal}), at $T=1.4K (\sim 0.93T_c)$.  The solid (dashed) line shows the real (imaginary) part of the Hall conductivity.  Compare with the full Hall conductivity at the same temperature in Fig.~\ref{SigmaOmegaTeq1pt4fig}.
}
\label{secondtermfig}
\end{minipage}
\hspace{1pc}
\begin{minipage}[t]{16pc}
\includegraphics[width=16pc]{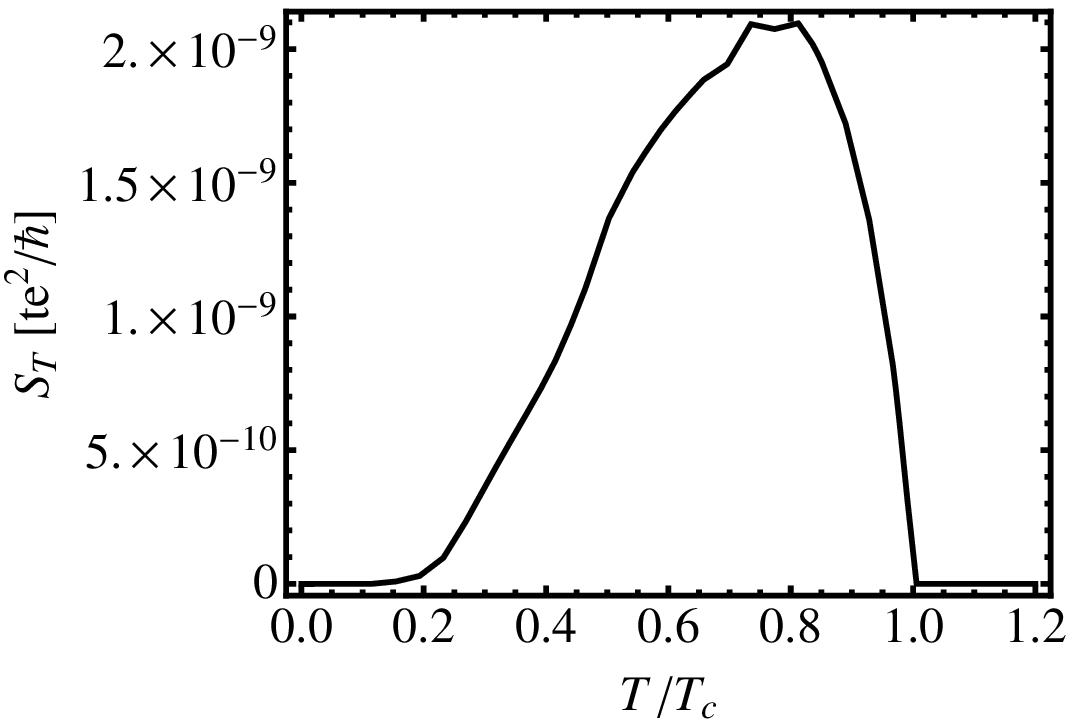}
\caption{Temperature dependence of the integrated spectral weight, $S_T\equiv -\int^{\infty}_0d\omega \mathrm{Im}\sigma^{T}_H(\omega)$, of the imaginary part of the contribution, $\sigma^{T}_H$, to the Hall conductivity from quasiparticle scattering (the second term in curly brackets in Eq.~(\ref{sigmareal})).}
	\label{secondtermTfig}
\end{minipage}
\end{figure}

The second term in Eq.~(\ref{sigmareal}) adds very little to the Hall conductivity.  For our parameters, we find this contribution ($\equiv \sigma^{T}_H$) to be always much smaller than the contribution coming from the first term and do not find the coherence features found by Wysoki\'{n}ski, Annett, and Gy\"{o}rffy at low frequencies, $\omega\sim \Delta_0$~\cite{Wysokinski12}.  Such features only arise in our calculations by artificially inflating the order parameter (and hence $T_c$ and crucially, the temperature) to energies on the order of several meVs.  Due to the thermal Fermi factor, $f(E_-)$,  which heavily suppresses all momenta except where $E_-\gtrsim  \Delta_0$ is small, $\sigma^T_H(\omega)$ will always be much smaller than $\sigma_H(\omega)$ unless $T/t^{\prime}$ and hence, $\Delta_0/t^{\prime}$, is appreciable. (Note, $E_+\gtrsim 2t^{\prime}$ and $f(E_+)\approx 0$.) 

Because of the Fermi factor, the largest contribution to $\sigma^T_H(\omega)$ will come when $\omega\sim E_+-E_-\sim 2t^{\prime}$, with $E_-\sim \Delta_0$ and $E_+\sim 2t^{\prime}$.  In Fig.~\ref{secondtermfig}, we plot the frequency-dependence of $\sigma_H^{T}(\omega)$ at $T\simeq 0.93T_c$ showing this strong frequency dependence~\cite{plotnote}.  To better understand the temperature dependence of this contribution, in Fig.~\ref{secondtermTfig}, we also plot the integrated spectral weight (defined to be positive)
\beq S_T\equiv -\int^{\infty}_0d\omega \mathrm{Im}\sigma^{T}_H(\omega) =  -\pi e^2\sum_{\bk}\frac{\epsilon_{12}\mathrm{Im}(\Delta^*_{11}\Delta_{22})\left(\delta\bv_{21}\times\bv_{12}\right)_z}{E_{+}E_{-}(E_{+}\!-\!E_{-})^2} \left[f(E_{+})\!-\!f(E_{-})\right]\label{sumrule}\eeq
of $\mathrm{Im}\sigma^{T}_H(\omega)$ in the clean limit $(\eta\to 0^+)$ as a function of temperature.  As expected, the low-temperature spectral weight from scattering $E_-$ BCS quasiparticles to $E_+$ is exponentially suppressed.

\section{Kerr angle at finite temperatures}
We now turn our attention to the Kerr angle, the quantity that has been measured in experiments~\cite{Xia06,Kapitulnik09} and which provides an indirect measure of the Hall conductivity.   The frequency-dependent  Kerr angle $\theta_K$ is given by  Eq.~(\ref{Kerr}). (See, for instance, Ref.~\cite{Lutchyn09}.)  In addition to depending on the frequency-dependent Hall conductivity, $\sigma_H(\omega)$, the Kerr angle also depends on the diagonal element of the optical conductivity tensor, $\sigma(\omega)$, through its dependence on the complex index of refraction $n(\omega)$:
\beq \alpha(\omega) = \frac{1}{n(n^2-1)},\label{alpha}\eeq
where $n(\omega)=\sqrt{\varepsilon_{ab}(\omega)}$ is equal to the square root of the component of the permeability tensor, \beq \varepsilon_{ab} = \varepsilon_{\infty}+(4\pi i/\omega)\sigma(\omega) ,\label{permeability}\eeq
 in the $ab$ plane.  Here, $\varepsilon_{\infty}$ is the background dielectric tensor.

   \begin{figure}
\begin{center}
\includegraphics[width=0.5\textwidth]{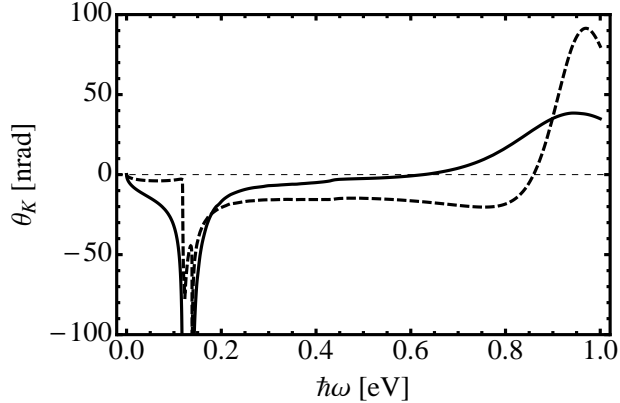}
\caption{Frequency dependence of the Kerr angle at $T=0$.  The solid line shows the Kerr angle for an inverse transport lifetime 0.4eV; the dashed line corresponds to an inverse transport lifetime 0.2eV.  The large difference between the two results shows the sensitivity of the Kerr angle to material parameters beyond the Hall conductivity.  Note the large resonant feature at $\omega \gtrsim 0.1$eV arising from the van-Hove singularity in the Hall conductivity (see Fig.~\ref{SigmaOmegaTeq0fig}).  
}
\label{Kerrfig}
\end{center}
\end{figure}
   
Having calculated the Hall conductivity, we can combine this calculation with experimental conductivity data to determine the frequency-dependent Kerr angle.  The conductivity in $\Sr$ seems to be well approximated by a generalized Drude expression with a frequency-dependent effective mass and transport scattering rate~\cite{Katsufuji96}.  For simplicity, however, we use a Drude model with constant transport scattering rate $\gamma$:
\beq \sigma(\omega) =- \frac{\omega^2_{pl}}{4\pi i(\omega + i\gamma)}\label{sigmaDrude} .\eeq
This is a reasonable approximation since the transport scattering rate varies relatively slowly over a large frequency range (see Fig.~(4b) in Ref.~\cite{Katsufuji96}), from approximately 2.5 eV$/\hbar$ at $\hbar\omega=0.1$eV to 4.5 eV/$\hbar$ at $\hbar\omega = 1$eV at low temperature. This variation is small in comparison with that exhibited by the Hall conductivity over the same range and we simply use the value at 0.8eV (the frequency at which the Kerr angle is measured in Ref.~\cite{Xia06}) to compute the frequency-dependent Kerr angle: $\gamma(0.8\mathrm{eV}) \sim 0.4$eV/$\hbar$.  

The other parameters needed to calculate the Kerr angle are also extracted from experiments on $\Sr$.  Following Ref.~\cite{Lutchyn09}  we take the interlayer spacing $d=6.8\AA$, $\varepsilon_{\infty}=10$, and $\omega_{pl}=2.9\mathrm{eV}$.  Using these, in  Fig.~\ref{Kerrfig} we plot the $T=0$ Kerr angle $\theta_K$ as a function of frequency using our calculated $T=0$ Hall conductivity (see Fig.~\ref{SigmaOmegaTeq0fig}).  The results are shown for two values of the transport scattering rate, the ``experimental value" of  $0.4$eV/$\hbar$ (extracted at $\omega=0.8$eV) and a somewhat smaller value of 0.2eV/$\hbar$. Using the former value, we find $\theta_K \simeq 17$nrads at $\hbar\omega = 0.8$eV, while the latter value gives $\theta_K  \simeq -20$nrads.  As one sees from the plot, in the vicinity of the plasma edge, the precise value of the Kerr angle is very sensitive to material parameters such as the transport scattering rate.  Setting the scattering rate to zero, as is sometimes done, is not a good approximation because of the significant inelastic scattering at the frequencies of interest.   

Over the frequency range of interest, the temperature dependence of the Kerr angle will track that of the Hall conductivity (see Figs.~\ref{ReSigmaTfig} and \ref{ImSigmaTfig}), which in turn is approximately proportional to the square of the order parameter.

\section{Discussion}
In this paper, we have calculated the anomalous Hall conductivity in a simple two-orbital model of a chiral superconductor, with an eye to understanding the Kerr effect in $\Sr$~\cite{Xia06}.  Although $\Sr$ is a three-orbital system, we have argued that the essential physics responsible for the anomalous Hall effect arises from only the Ru $d_{xz}$ and $d_{yz}$ orbitals, since these are the only orbitals that are  optically coupled.  This fact gives an important experimental handle in understanding \emph{where} superconductivity arises in $\Sr$: In contrast to the impurity-scattering extrinsic contribution to the Hall conductivity~\cite{Goryo08,Lutchyn09}, there must be significant pairing on the $\alpha$ and $\beta$ bands in order to have an intrinsic anomalous Hall conductivity.   The ``smoking gun" for such superconductivity would be a large resonant feature in the Hall conductivity at a frequency close to the twice the inter-orbital hopping scale $t^{\prime}\sim 0.04$eV, as clearly seen in Figs.~\ref{SigmaOmegaTeq0fig} and \ref{SigmaOmegaTeq1pt4fig} (also in Fig.~\ref{Kerrfig} for the Kerr angle).  This feature is robust both to nonzero temperature as well as impurity broadening effects since both these energy scales are on the order of $\Delta_0$,  much smaller than $2t^{\prime}$.  

To make contact with the Kerr rotation experiment~\cite{Xia06}, using our two-orbital model in conjunction with parameters appropriate for $\Sr$, we have arrived at the result $\theta_K(0.8\mathrm{eV})\simeq 17$nrads for the Kerr angle at $T=0$.  This is of the same order of magnitude---but smaller than---the experimental value of $\sim 65$nrads~\cite{Xia06}. 

Although the inclusion of the $d_{xy}$ orbital would not lead to any qualitatively new physics, it can lead to a quantitative change to this prediction.   We also note that the $d_{xz}/d_{yz}$ hopping parameters used in our calculations were obtained from a microscopic \emph{three}-orbital model~\cite{Rozbicki11} including spin-orbit coupling (SOC), which we have also excluded.  As we showed in Ref.~\cite{Taylor12}, SOC does not lead to any Hall effect unless there is a Zeeman splitting $\mu_{\uparrow}\neq \mu_{\downarrow}$ between the two spin species.  Its only effect otherwise is to renormalize the dispersions that enter Eq.~(\ref{sigmareal}).  

In comparison to our earlier work~\cite{Taylor12}, where we predicted a $T=0$ Kerr angle of $50$nrads using a two-orbital model without next-nearest neighbour hopping, the present work produces a smaller Kerr angle, resulting from a smaller prediction for the Hall conductivity.  This difference is due almost entirely to the form for the order parameter given in Eq.~(\ref{ansatz}), appropriate for next-nearest neighbour hopping, an effect not included in our earlier calculations.  The additional cosine factors in the order parameters lead to a significant reduction in the Hall conductivity since the integrand in Eq.~(\ref{sigmareal}) is dominated by a small region in the vicinity of $k_x\sim k_y\lesssim |\bk_0| \equiv \cos^{-1}(-\mu/2t)$ (see Eq.~(\ref{QPmin})).  This reduction is offset somewhat by the larger value of $\Delta_0(T=0)\simeq 2.45T_c$ (where $T_c=1.5$K) used in the present work as a result of solving the BCS gap equation appropriate for our model, instead of using the $s$-wave BCS result $\Delta_0(T=0)\simeq  1.76T_c$

Although the Kerr angle is additionally very sensitive to the low-temperature optical properties of $\Sr$, and more information about these are needed to reliably make contact with experiment, there are a few ways in which experiments should be able to distinguish between the two contributions -- extrinsic~\cite{Goryo08,Lutchyn09} and intrinsic -- to the Hall conductivity identified by theory.  First, as we have noted in this paper, the intrinsic mechanism will exhibit a strong frequency dependence in the vicinity of twice the inter-orbital hopping, around 40 meVs.  Unfortunately, this seems to be outside the range of the current experimental setup at Stanford~\cite{Xia06,Kapitulnik09}.  Secondly, experiments could probe the dependence of the Kerr angle on the impurity concentration.  

To a good approximation, the intrinsic mechanism discussed here leads to a Hall conductivity that depends on the impurity concentration only through its approximately quadratic dependence on the order parameter, Eq.~(\ref{propid}).  In turn, the impurity dependence of $T_c$, and hence, $\Delta_0$, has been  studied~\cite{Mackenzie98a,Mackenzie98b}.  In contrast, the extrinsic contribution to the Hall conductivity~\cite{Goryo08,Lutchyn09} will exhibit an additional, stronger dependence on the impurity concentration.  

In summary, a more detailed understanding of the anomalous Hall conductivity in $\Sr$, either by probing its frequency dependence or its dependence on the concentration of non-magnetic impurities, would contribute significantly to understanding this material.  In particular, the Hall conductivity is sensitive to the way in which the different bands crossing the Fermi energy of $\Sr$ enter into the chiral $p$-wave order parameter.   We hope that the present work stimulates further experimental studies of the anomalous Hall effect in $\Sr$ as well as in any other materials that might exhibit chiral superconductivity~\cite{Nandkishore12}.  

\ack 

We thank Steve Kivelson, Sri Raghu, and Ronny Thomale for useful
discussions.  This work was supported by NSERC and the Canadian Institute for Advanced Research.


\begin{thebibliography}{9}
\bibitem{note} Although we suppress spin in our model, the Hall conductivity receives equal contributions from both spins.  The current operator includes contributions from both particle and hole contributions, mimicking spin counting.  Hence, the Hall conductivity calculated from this spin operator correctly accounts for two spins.
\end{thebibliography}
\end{document}